\documentclass[
    ,final            
  ]
  {aipproc}

\layoutstyle{8x11double}

\begin{document}

\title{Small scale WIMP physics}


\author{Anne M.~Green}{
  address={Physics and Astronomy, University of Sheffield, Sheffield
    S3 7RH, UK}
}
\author{Stefan Hofman}{
  address={Perimeter Institute for Theoretical Physics, Waterloo, Ontario, N2L2Y5, Canada}
}

\author{Dominik J.~Schwarz}{
  address={Fakult\"at f\"ur Physik, Universit\"at Bielefeld, 
  Postfach 100131, 33501 Bielefeld, Germany}
}

\begin{abstract} 

The dark matter distribution on small scales may depend on the the
properties of the first generation of dark matter halos to form, which
is in turn determined by the microphysics of the dark matter
particles.  We overview the microphysics of WIMPs and our calculations
of the collisional damping and free streaming scales.  We then plot
the resulting density perturbation power spectrum and red-shift at which 
typical halos form, taking into
account the effect of uncertainties in the WIMP properties (mass and
interaction channel) and the primordial power spectrum.  Finally we
review recent developments regarding the properties and fate of the first
WIMPy halos.

\end{abstract}

\maketitle


\section{Why?}

Diverse cosmological observations indicate that the Universe contains
a significant amount of non-baryonic cold dark matter (CDM).
Weakly interacting massive particles (WIMPs) are an attractive CDM
candidate, as a stable relic from the electroweak scale
generically has an interesting present day density, $\Omega_{\rm cdm}
\sim {\cal O}(1)$.

According to standard lore the distribution of CDM is independent of
its nature (WIMPs, axions, or something more exotic). This is true on
large (super-kpc) scales, but may not be the case on small (sub-pc)
scales. Furthermore the signals expected in WIMP detection experiments
depend on the dark matter distribution.  WIMP direct
detection experiments (which aim to detect WIMPs in the lab via their
elastic scattering off detector
nuclei) probe the dark matter distribution on sub-milli-pc scales. WIMP
indirect detection involves searching for the products ($\gamma$-rays,
antiprotons and neutrinos) of WIMP annihilation and, since this is a
two-body process, the event rates are proportional to the density
squared and are hence  enhanced by substructure. Reliable
predictions for the expected signals therefore require an
understanding of the distribution of dark matter on small (sub-galactic)
scales.

In CDM cosmologies structure forms hierarchically; small halos form
first with larger halos forming via mergers and accretion.  The
density perturbations on very small scales, and hence the properties
of the first generation of structures to form, depend on the
micro-physics of the CDM and the present day density distribution may
retain traces of these first structures.

\section{WIMP micro-physics}

In the early Universe WIMPs are kept in thermal equilibrium with the
radiation component via chemical interactions.  As the Universe
expands the WIMP density decreases, these interactions cease and the
WIMPs chemically decouple or `freeze-out'. After this the WIMP
comoving number density remains constant, however WIMPs continue to
interact kinetically, via elastic scattering, with the radiation
component and are kept in local thermal equilibrium. Eventually the
WIMP density becomes so low that elastic scattering ceases and the
WIMPs kinetically decouple~\cite{ssw,ckz,hss}. The average
momentum exchanged per collision is small however so the relaxation
timescale, which characterises the time at which WIMPs kinetically
decouple, is significantly larger than the elastic scattering
timescale~\cite{ssw,hss}. The kinetic decoupling
temperature was first estimated in Ref.~\cite{ssw} while
Ref.~\cite{ckz} calculated the last scattering temperature for
supersymmetric WIMPs. While chemical decoupling happens at a
temperature $T_{\rm cd} \sim {\cal O} (10$ GeV), kinetic decoupling is
delayed by the large entropy of the hot Universe and takes place at a
temperature $T_{\rm kd} \sim {\cal O} (10$ MeV).  This is generic for
any WIMP that is of cosmological relevance today.

Elastic scattering processes around $T_{\rm kd}$ lead to viscous
coupling of the WIMP and radiation fluids, resulting in
collisional damping of WIMP perturbations due to bulk and shear
viscosity~\cite{hss}.  The collisional damping scale was first
derived in Ref.~\cite{hss} by two of the present authors,
and we presented a more intuitive, but less rigorous, derivation
using the linearised Navier-Stokes equation in Ref.~\cite{ghs2}.
The net result is exponential damping with a characteristic wave-number
\begin{equation}
k_{\rm d} =
\frac{3.76\times 10^7}{\rm Mpc}\left(\frac{m}{100 \; {\rm GeV}}\right)^{1/2}
\left(\frac{T_{\rm kd}}{30 \; {\rm MeV}}\right)^{1/2} \,,
\end{equation}
where $m$ is the WIMP mass.

One caveat is that our treatment is restricted to sub-horizon scales.
Near the horizon there are additional effects, which have recently
been studied numerically by Loeb and Zaldarriaga~\cite{LZ}. Their
results confirm that the essential features of the subhorizon damping
are captured by our calculation, but the additional
terms play an important role in determining the precise position of
the maximum of the CDM power spectrum. We agree with their conclusion
that an accurate calculation (better than $10\%$ accuracy) must be
based on a numerical computation including all of the terms.

After kinetic decoupling the WIMPs free-stream and their distribution
function is governed by the collisionless Boltzmann equation.  We
study the subsequent additional damping 
by considering small perturbations away from
the local thermal equilibrium solution, taking into account the
perturbations present at kinetic decoupling. The characteristic
free-streaming scale becomes constant after matter-radiation equality
and is given by
\begin{equation}
k_{\rm fs} =
\frac{1.70 \times 10^6}{\rm Mpc}
\frac{(m/100 \; {\rm GeV})^{1/2} (T_{\rm kd}/30 \; {\rm MeV})^{1/2}}
{1+ {\rm ln}(T_{\rm kd}/30 \; {\rm MeV})/19.2} \,.
\end{equation}
The free streaming leads to exponential
damping of the WIMP density contrast again but 
with an additional polynomial pre-factor~\cite{ghs,ghs2}. 
Since $k_{\rm fs} < k_{\rm d}$, the cut-off in the power spectrum
is determined by the free streaming scale $k_{\rm fs}$.

\begin{table}

\label{WIMPtable}
\begin{tabular}{ccccccc}
\hline
Ref. & $l$& $m$ (GeV)
& $T_{\rm cd}$ (GeV) & $T_{\rm kd}$ (Mev)& $k_{\rm d} \, ({\rm pc}^{-1})$
&  $k_{\rm fs}  \,({\rm pc}^{-1})$\\
\hline
A &0&  100 & 3.6 & 1.6 & 14 & 0.42 \\
B & 1 & 50 & 1.9 & 21 & 39 & 0.94 \\
C & 1 & 100 & 3.7 & 25 & 61 & 1.5 \\
D & 1 & 500 & 17 & 37 & 180 & 4.0 \\
\hline
\end{tabular}
\caption{Benchmark WIMP models.}
\end{table}

We focus on four benchmark models which span the range of most plausible
WIMP properties and have present day densities in accordance with the
WMAP measurement of the cold dark matter density. The details of these
benchmark models, including the values of $k_{\rm d}$ and $k_{\rm
fs}$, are tabulated in Table 1. $l$ parameterises the temperature
dependence of the thermally averaged elastic scattering cross section:
$\langle \sigma_{\rm el}\rangle = \sigma^{\rm el}_0 (T/m)^{1+l}$. In
the Standard Model, elastic scattering between a light fermion and a
heavy fermion is mediated by ${\rm Z}^0$ exchange and $l=0$. Other
channels may occur however. In supersymmetric extensions of the
Standard Model, where the lightest neutralino is a WIMP candidate,
sfermion exchange occurs (and if the neutralino is a gaugino, ${\rm
Z}^0$ exchange is suppressed), in which case $l=1$.  Models B and C
are very close to the bino-like neutralino case considered 
in Ref.~\cite{ghs}. Models A and D show that there is more spread in
the predicted damping scale if the assumption of a bino-like
WIMP is dropped.

\begin{figure}
  \includegraphics[height=.25\textheight]{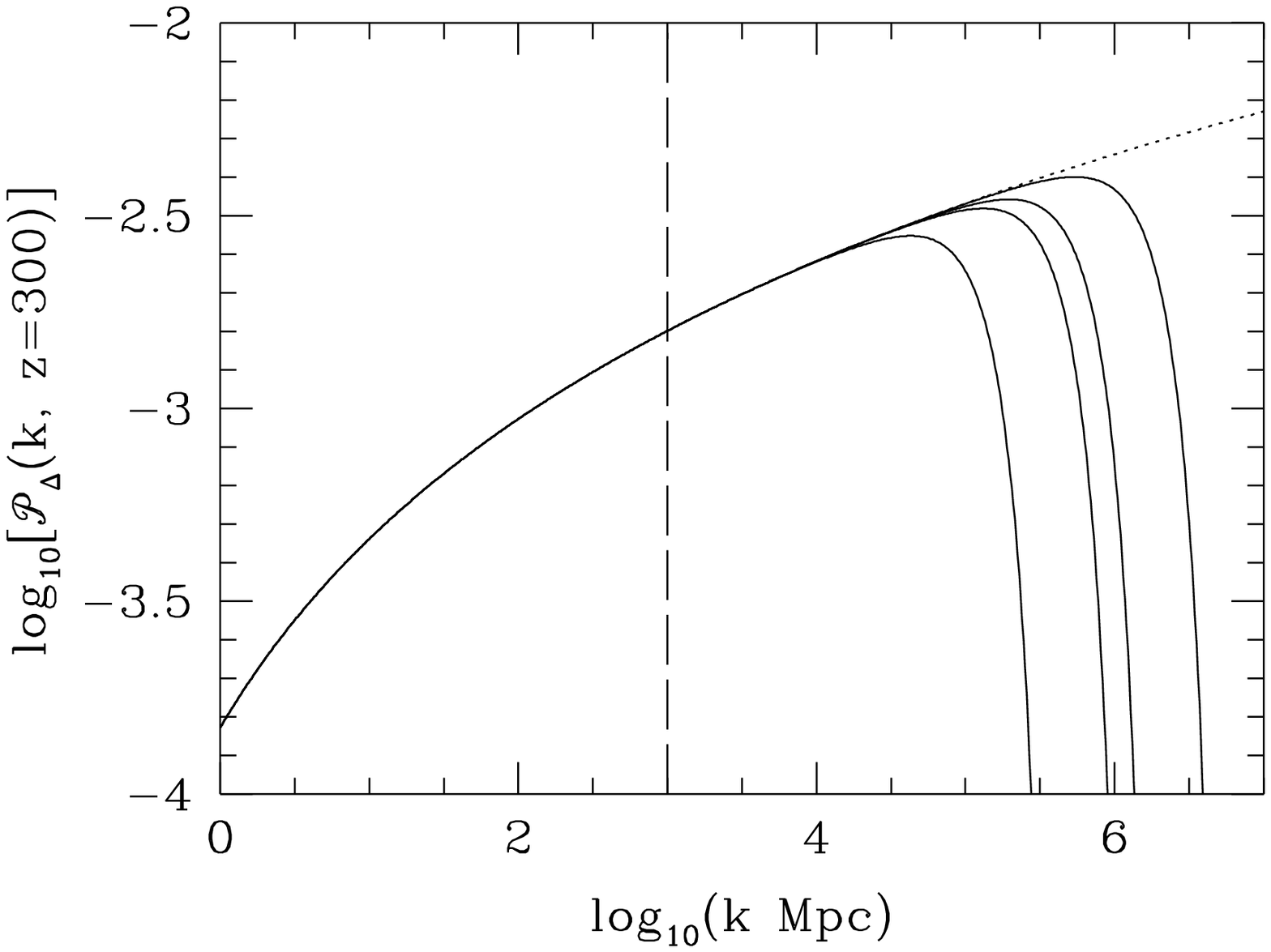}
\includegraphics[height=.25\textheight]{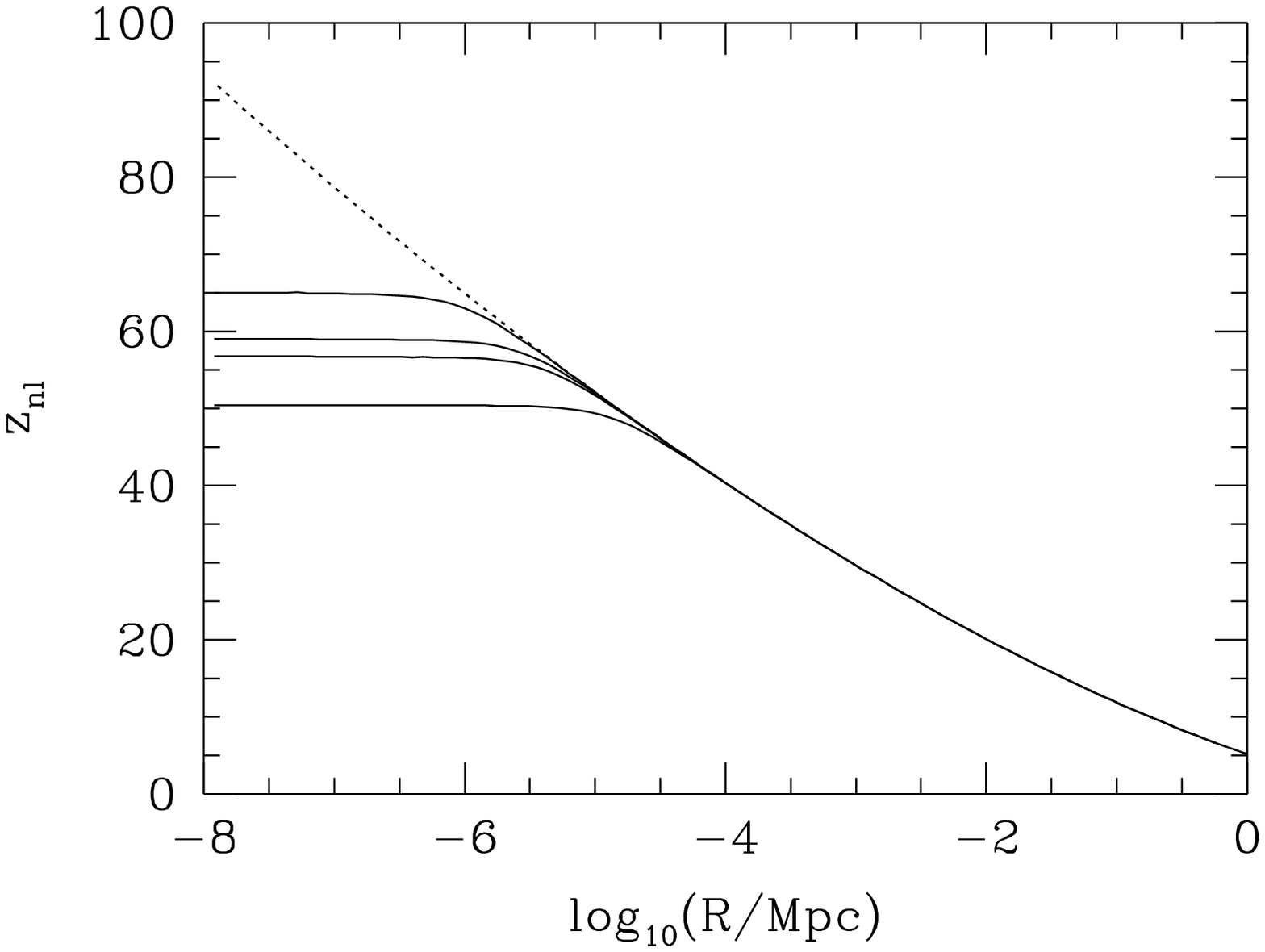}
  \caption{
The processed power spectrum at $z=300$, ${\cal
 P}_{\Delta}(k,z=300)$, (left)
  and the red-shift at which $1-\sigma$ fluctuations go non-linear, 
$z_{\rm nl}$, (right) for a WMAP normalised scale-invariant primordial
power spectrum and from left to right/bottom to top
a Dirac WIMP with $m=100$ GeV and Majorana WIMPs with $m=50, 100, 500$ GeV. 
Without the effects of collisional damping
and free streaming these quantities would be given by the dotted line.
The vertical dashed line denotes $k_{\rm b}$, the wavenumber below which
baryons follow CDM. Our approximations are optimised for $k>k_{\rm b}$.}
\end{figure}

\section{Power spectrum and $z_{\rm nl}$}

The (processed) power spectrum depends on two more ingredients: the
gravitational growth of perturbations and the primordial power
spectrum.  We calculate the transfer function, which encodes the
gravitational growth, by solving the equations which govern the
evolution of perturbations in matter and radiation fluids during two
overlapping epochs, radiation domination and close to matter-radiation
equality, and then matching these solutions together (see
Ref.~\cite{ghs2} for further details). We neglect
perturbations in the baryon fluid. This approximation is valid on
small scales $k>k_{\rm b} \sim 10^{3} \, {\rm Mpc}^{-1}$ for $z> z_{\rm
b} \sim 150$. We have verified that our analytic expressions are
accurate at the 10$\%$ level by comparing them with the output of the
COSMICS package. This is sufficient given current uncertainties in the
cosmological parameters and the primordial power spectrum.

The resulting linear power
spectrum, defined as
${\cal P}_{\Delta}(k,z) = k^3/(2\pi^2) \langle|\Delta(k,z)|^2\rangle$,
at $z=300$ (before the onset of non-linear structure formation)
is plotted in the left panel in Fig.1 for our benchmark
WIMP models and a WMAP normalised scale-invariant primordial
power spectrum. The characteristic free-streaming wave-number is
smallest for model A (Dirac WIMP with l=0) and increases with
increasing mass for the Majorana WIMPs (models B-D). This is clearly
reflected in the position of the cut-off in the power spectrum in Fig. 1.

The cut-off in the WIMP power spectrum sets the scale
of the first halos to form. We estimate the redshift at which 
typical (1-$\sigma$) fluctuations on
comoving scale $R$ go nonlinear via $\sigma(R, z_{\rm nl}) = 1$, 
where $\sigma(R,z)$ is the mass variance defined by
\begin{equation}
\sigma^2(R, z) = \int_{0}^{\infty} W^2(kR) {\cal P}_{\Delta}(k,z)
        \frac{{\rm d} k}{k} \,,
\end{equation}
where $W(kR)$ is the Fourier transform of the window function, 
which we take to be a top hat, divided by its volume. 

The right panel of Fig. 1 shows $z_{\rm nl}$ as a function of scale
$R$.  The cut-off in the processed power spectrum at $k \sim 10^{6} \,
{\rm Mpc}^{-1}$ leads to a plateau with $z_{\rm nl}= z_{\rm nl}^{\rm
max}$ at $R < R_{\rm min} = {\cal O}(1)$ pc.  We emphasise that
$z_{\rm nl}^{\rm max}$ is the redshift at which hierarchical structure
formation starts at a typical place in the universe, the very first
WIMPy halos will form significantly earlier from rare large
fluctuations.  The order of magnitude variation in $k_{\rm fs}$ leads
to a similar variation in $R_{\rm min}$ and also (because of the
dependence of the amplitude of the peak of the power spectrum on the
cut-off scale) a range of values $z_{\rm nl}^{\rm max} \approx 50$ to
$65$.

\begin{figure}
  \includegraphics[height=.25\textheight]{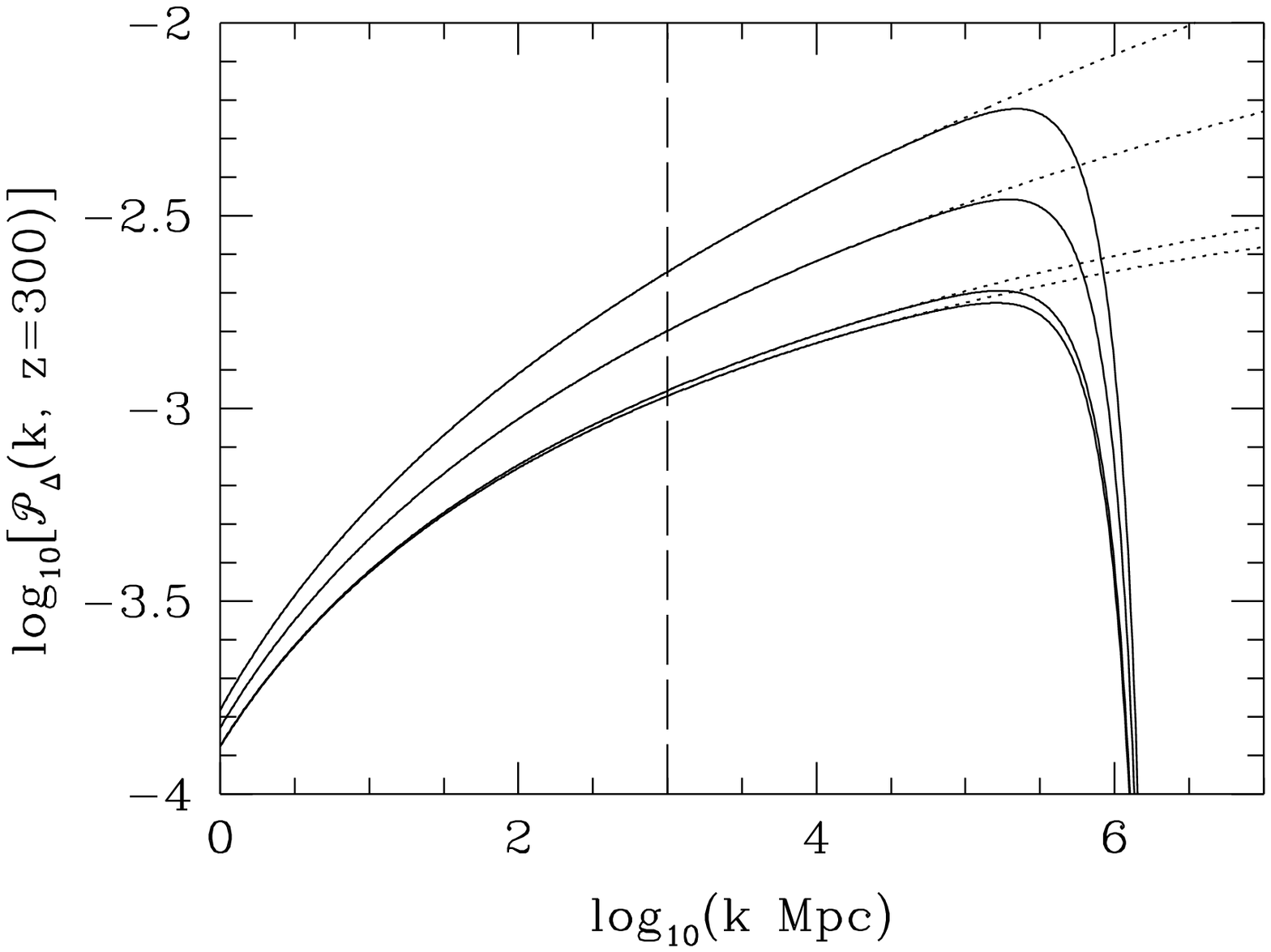}
\includegraphics[height=.25\textheight]{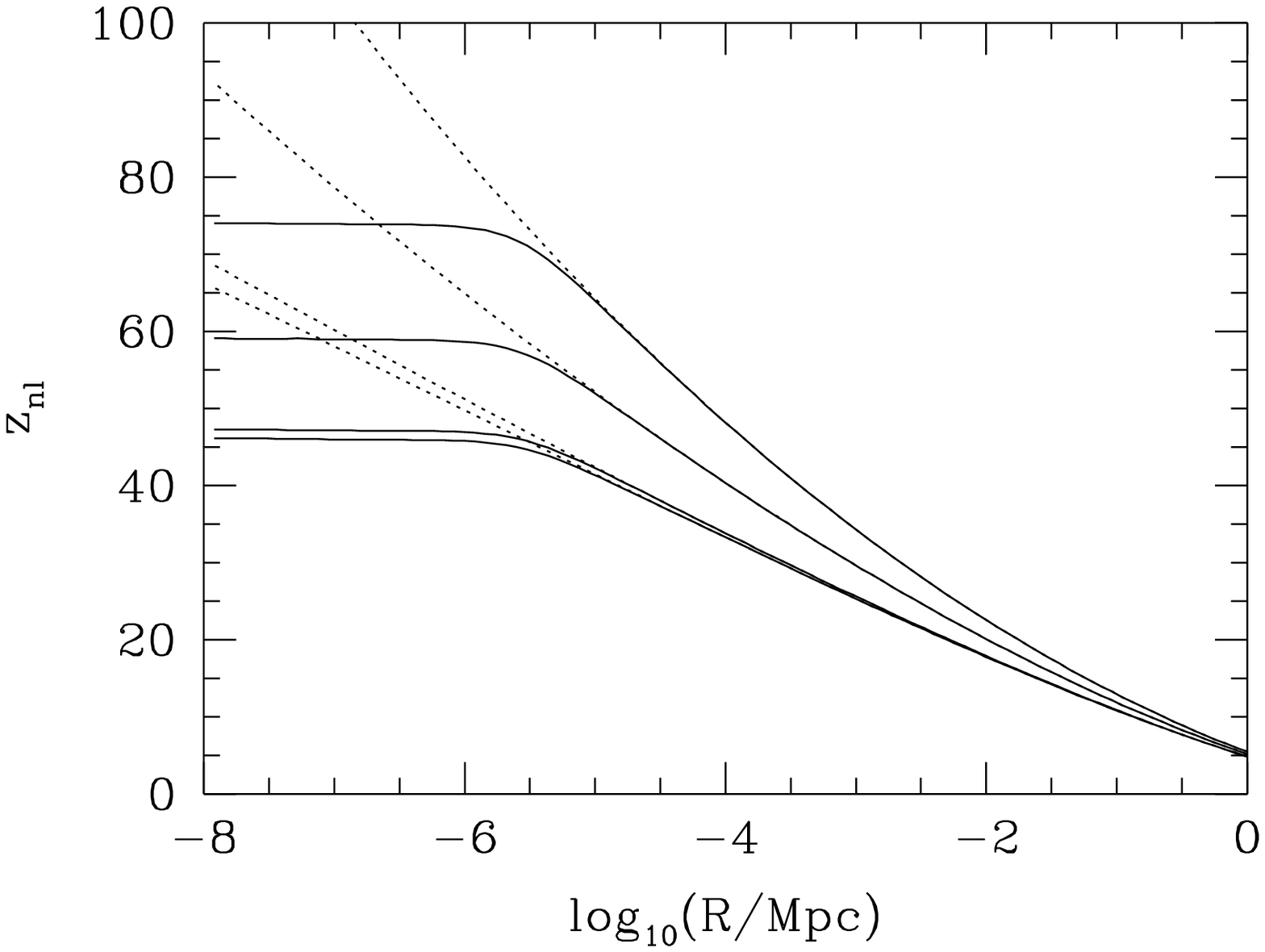}
  \caption{As Fig. 1 for WIMP benchmark model C 
(a bino like WIMP with $m=100$ GeV) and from top to 
bottom false vacuum dominated hybrid inflation ($n=1.036, \alpha=0$),
a scale invariant primordial power spectrum ($n=1, \alpha=0$), 
power law inflation
($n=0.964, \alpha=0$) and $m^2 \phi^2$ chaotic inflation 
($n=0.964, \alpha=-0.0006$).
 }
\end{figure}

We now turn our attention to the effect of the uncertainty in
the primordial power spectrum.
On the scales probed by the CMB [${\cal O} (0.01-0.1) \, {\rm Mpc}^{-1}$]
the primordial power spectrum is close to scale
invariant. The free-streaming scale $k_{\rm fs} \sim
10^{6} \, {\rm Mpc}^{-1}$ is seven orders of magnitude smaller and therefore
even a very small scale dependence of the power spectrum could
significantly change the amplitude of the power spectrum at the
cut-off scale, and hence the red-shift at which the first WIMP halos form.
The most commonly used parameterisation of the 
primordial power spectrum is
\begin{equation}
\label{prim1}
{\cal P}(k)={\cal P}(k_{0}) \left(
          \frac{k}{k_{0}} \right)^{n(k_{0})-1 + \frac{1}{2} 
     \alpha(k_{0}) {\rm ln}(k/k_{0})} \,,
\end{equation}
where $\alpha(k)={\rm d} n/ {\rm d} k$. The spectral index $n$ and 
its running $\alpha$ depend on the inflationary potential.

To assess the effects of possible scale dependence of the primordial
power spectrum we consider three benchmark inflation models which span
the range of possible power spectra for simple inflation models:
$V=m^{2}\phi^{2}$ chaotic inflation, with $N=55$ e-foldings of
inflation so that $n-1=-0.036$ and $\alpha=-6 \times 10^{-4}$,
power law inflation ($a \propto t^p$), with $p=55.4$ so that
$n-1=-0.036$, to match the chaotic inflation model, but in this case
$\alpha=0$ and false vacuum dominated hybrid inflation with the
largest spectral index allowed by the WMAP data, $n-1=0.036$ and
$\alpha=0$. For further details see Sec.~7 of Ref.~\cite{ghs2}.

The amplitude of the primordial power spectra on the free-streaming
scale $\sim 10^{6} {\rm Mpc}^{-1}$ varies by a factor of $\sim 2.5$
(equivalently the amplitude of the fluctuations varies by $\sim
\sqrt{2.5} \sim 1.6$).
The resulting processed power spectra and the red-shift at which
scales go non-linear are plotted in Fig.2 for a Majorana WIMP with
$m=100 \, {\rm GeV}$. The variation in the amplitude of the primordial power
spectra at the cut-off scale translates directly into a variation in
the peak amplitude of the processed power spectra and hence 
$z_{\rm nl}^{\rm max}$.


\section{Fate of the first halos}

We estimate the size and mass of the first generation of typical
subhalos that form at $z_{\rm nl}^{\max}$ using the spherical collapse
model~\cite{ghs,ghs2}.  The mean CDM mass within a
sphere of comoving radius R is $M(R)= 1.6 \times 10^{-7} M_{\odot}(
R/{\rm pc})^3$, with over-densities having twice this mass.
The present day radius, after turn-around and violent relaxation, would
be of order tens of milli-pc, comparable to the size of the solar
system, and smaller.

Numerical simulations are required to study the detailed properties,
and fate, of the first generation of WIMPy halos to form. Diemand et
al.~\cite{dms} have recently published the results of the first such
simulations to be carried out.  They use a multi-scale technique to
study the non-linear collapse of a small, $(60 \,\, {\rm comoving \,
pc})^3$, average density region starting at $z=350$.  The first
non-linear structures form at $z\sim 60$ and have $M \sim 10^{-6}
M_{\odot}$, in good agreement with our analytic calculations. The
simulations have to be stopped at $z\sim 25$ when the high resolution
region begins to merge with the surrounding lower resolution regions.

These mini-halos are subsequently subject to a number of dynamical
processes which could destroy them. Matter will be tidally stripped
from the outer regions if the gravitational force of the parent halo
exceeds that of the sub-halo. Diemand et al. estimate that the mini
halos should not be tidally destroyed outside the inner few parsecs of
the Milky Way~\cite{dms}, in contrast to earlier analytic calculations by
Berezinsky et al.~\cite{beren}. The mini-halos can also be destroyed
via interactions with compact objects. Zhao et al. argue that
encounters with stars destroy all of the mini halos in the solar
neighbourhood~\cite{zhao}, Moore et al.  however
disagree~\cite{moore}. Further numerical simulations of the dynamical
evolution and interactions of mini-halos are probably required to
resolve these issues.

\section{Summary} 

WIMP direct and indirect detection both probe the dark matter
distribution on small scales, which depends on the properties, and
hence fate, of the first generation of WIMP halos to form. Collisional
damping and free-streaming erase density perturbations on small scales
($k>10^{6} \, {\rm Mpc}^{-1}$) and set the scale of the first generation
of WIMP halos to form ($M \sim 10^{-6} \, M_{\odot}$). The fate of these
halos, specifically whether or not a significant fraction survive to the
present day within the Milky Way, remains an open question.








\IfFileExists{\jobname.bbl}{}
 {\typeout{}
  \typeout{******************************************}
  \typeout{** Please run "bibtex \jobname" to optain}
  \typeout{** the bibliography and then re-run LaTeX}
  \typeout{** twice to fix the references!}
  \typeout{******************************************}
  \typeout{}
 }

\end{document}